# Synergies between AI Computing and Power Systems: Metrics, Scheduling, and Resilience

Farzaneh Pourahmadi, Olivier Corradi, Pierre Pinson

## 1. Introduction

While the idea of "Green AI" is gaining momentum, a vital piece of the puzzle remains largely unaddressed: the intersection of AI computing and power systems. Despite their vast energy footprints, major data centers run by companies like Google, Microsoft, and Amazon (among others) still operate without clear sustainability guidelines or regulatory frameworks. The potential environmental toll remains largely hidden from public view.

To put this into perspective, training a single large AI model can emit 300 tons of $CO_2$ which is the equivalent of 125 direct flights between New York and Beijing[1]. This is not an outlier. Global energy demand from data centers grew from 194 terawatt-hours (TWh) in 2010 to 460 TWh by 2022, with projections suggesting it could rise to as much as 1,050 TWh by 2026 [1]. If left unchecked, this trajectory risks undermining global climate goals. Part of the problem is that the true costs of large-scale AI remain opaque to most people. The widespread belief that building an AI model is a purely digital, low-impact process obscures the reality of its physical footprint. This misconception is fueled *by the lack of standardized and transparent metrics* that connect compute tasks with environmental footprints.

To make this concrete, Google's own production measurements for Gemini report that a typical text prompt uses about 0.24 Wh, roughly the energy of watching TV for about nine seconds, which emits about 0.03 g $CO_2$e. Google also cites substantial year-over-year efficiency gains (up to 33× lower energy and 44× lower carbon per median prompt) from hardware, software, and data-center improvements [2]. Although the per-query impact is small, at the scale of billions of prompts it becomes material that underscores the need for time- and location-aware carbon signals rather than static averages.

Researchers and companies alike have begun publishing footprint, statistics, and building calculators, which is a step forward. Yet, most tools still overlook key variables like when energy is consumed, the flexibility of training schedules, the location of data centers within the power grid, and the real-time carbon intensity of electricity sources. Without these factors, carbon accounting remains incomplete. Moreover, current studies tend to focus narrowly on deep learning and large language models, making it difficult to generalize findings across the diverse landscape of AI. *What is needed is a broader, more dynamic perspective that views AI not just as a passive consumer of energy, but as an active participant in the digital complex power system.*

---

[1] Calculation using https://co2.myclimate.org

By *(i)* intelligently planning when and where data centers are located in the system and *(ii)* smartly scheduling how AI workloads run, we could significantly reduce emissions without sacrificing performance. Yet today, there are no formal incentives, signals, or policies to encourage this kind of climate-aware computing. In short, a sustainable AI future demands deeper integration between the worlds of AI computing and power systems. Until we bridge that gap, we will lack the metrics, coordination mechanisms, and planning frameworks needed to guide AI development in a way that truly supports the green transition.

> Without aligning the growing demands of AI computing with the capabilities of power systems, we risk letting technological progress quietly accelerate environmental harm, rather than contribute to the green transition.

In this paper, we first clarify the concepts of "Green AI" versus "Frugal AI," positioning frugality as efficiency by design and Green AI as transparency and accountability. We then argue that these approaches, while complementary, are insufficient without a shared quantitative foundation that links AI computing to power system contexts. This motivates the development of standardized carbon metrics as a bridge between algorithmic decisions and their physical consequences.

Second, we review existing carbon accounting methods and motivate their evolution into dynamic, privacy-aware, and fair coordination signals that can align many local computing decisions with system-level emissions reduction.

Third, we embed these coordination signals into scheduling and planning frameworks, demonstrating two complementary architectures: (i) an iterative signal–response loop for day-ahead, intra-day, and real-time operation, and (ii) an integrated optimization that learns and encodes flexible-load behavior for long-term planning.

Finally, we show that the same coordination stack naturally extends to resilience, where signals can pivot from "emissions-first" to "stability-first," enabling targeted relief, faster restoration, and an orderly return to green operation. Here, resilience refers to the capability of coupled AI-computing and power-system infrastructures to absorb shocks, continue essential functions, and recover rapidly under stress conditions such as outages or extreme weather. Figure 1 illustrates and summarizes the coupled system we analyze and discuss in the paper. The operator estimates emissions and publishes time- and location-aware carbon coordination signals, while data centers respond through demand response, i.e., flexible adjustment of electricity use and workload scheduling in response to system signals, closing the loop closing the loop to lower emissions and bolster resilience.

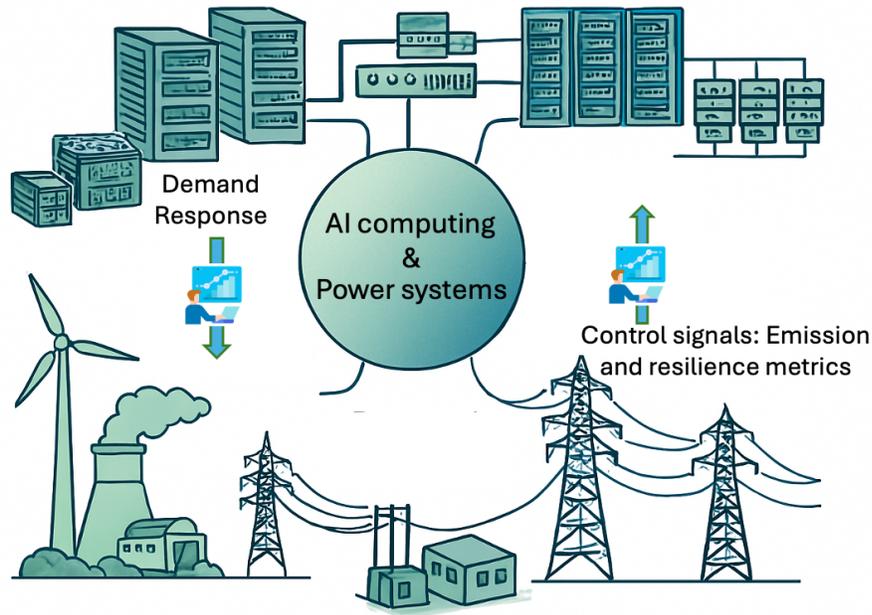

**Figure 1.** AI computing and power systems as a coupled system. The operator estimates emissions and publishes carbon coordination signals, and data centers respond through demand response, closing the loop between AI computing and the power system for decarbonization and resilience.

## 2. Green AI and Frugal AI

The environmental impact of AI is no longer just a side note, while it is becoming central to how we think about responsible technological progress. As AI models grow more complex, so does the need to assess and manage the computational workload required for training and inference. Yet research into the resource demands of AI, especially in relation to the broader digital power system, is still in its early stages. In recent literature, various terms like Green AI and Frugal AI are often used interchangeably. However, these terms describe different strategies, and they are rarely defined in any standardized way. Clarifying their meanings is crucial to understanding how we can design AI systems that are not just powerful, but also aligned with environmental goals.

While both Green AI and Frugal AI aim to reduce the environmental impact of AI, they differ in scope and intent. Green AI focuses on measurement, transparency, and accountability. It encourages researchers and developers to report energy use and carbon emissions associated with training and deploying models. The goal is to make the environmental cost of AI visible, with the hope that transparency will drive better decisions such as choosing more efficient models or cleaner infrastructure. In this sense, *Green AI is reactive*: it helps us see and respond to environmental impact after it has occurred. *Frugal AI, by contrast, is proactive.* It aims to minimize the use of resources from the outset by designing efficient algorithms, models, and workflows. Rather than simply measuring emissions, frugal AI asks how to achieve similar performance with fewer inputs whether in terms of data, computation, energy, or time. This concept is based on the principle of computational sufficiency, emphasizing that models should use only as much computation as necessary to achieve their purpose.

Frugal AI methods include well-known techniques like model pruning, quantization, and knowledge distillation, all of which reduce model size and complexity while maintaining acceptable accuracy. These approaches are particularly valuable for edge or embedded systems, where energy and memory are constrained. At the hardware level, the use of AI accelerators such as GPUs and TPUs, has also improved the energy efficiency of training and inference, particularly when paired with software optimizations. Other strategies like feature selection and data selection help streamline datasets by removing redundancy, ensuring models are trained only on the most informative inputs.

Despite these advances, critical challenges remain. Most Frugal AI methods are model-agnostic and do not provide clear performance guarantees. For instance, while pruning or distillation might reduce model size, there is no straightforward way to determine how often a model should be retrained, or which data points are most cost-effective for a given downstream task. These gaps raise important questions: When is it worth retraining a model? Which data and features are worth keeping? How can performance be maintained over time with limited resources?

This is analogous to employing large-scale AI models for routine or low-complexity tasks that could be handled efficiently by simpler algorithms. Smarter systems need to match computational effort with task importance and contextual constraints. Therefore, from an algorithmic perspective, Frugal AI should move beyond generic efficiency techniques and become more application-specific. Not all tasks require the same level of precision, speed, or model complexity, so AI systems should be tailored to the demands of their intended use. Designing task-aware, resource-conscious models can help avoid unnecessary computation and improve overall efficiency. For example, a lightweight model may be sufficient for routine monitoring tasks, while more complex architectures can be reserved for critical decision-making, ensuring that computational resources are used where (and when) they matter most.

From a systems perspective, even the best-designed frugal algorithms often ignore a crucial systemic factor: the power grid within which AI operates. Both Green and Frugal AI tend to treat electricity use as a static quantity, when in fact, the carbon intensity of electricity varies dramatically depending on, e.g., location, time of day, and grid conditions. Such coordination transforms AI from a passive energy consumer into a flexible, grid-aware participant. Integrating green AI and frugal AI with power system operations through real-time carbon signals, demand-response mechanisms, and energy pricing, offers a powerful lever for sustainable digital infrastructure. Together, these perspectives suggest a more holistic approach: building AI that is efficient by design and strategically deployed in time and space. Only through this combined lens can we unlock the full potential of AI to serve both innovation and the energy transition.

In summary, while Green AI and Frugal AI address transparency and efficiency at the algorithmic level, both require a common quantitative foundation that links computational choices to system-wide environmental impact. Standardized carbon-intensity metrics provide this foundation and form the basis for coordinating AI computing with power-system operation.

# 3. Green Metrics and Signals

From a power systems perspective, there is growing interest among individuals, companies, and organizations in understanding and ultimately reducing the carbon emissions associated with their electricity use in real time. This interest spans multiple sectors, including data centers, hydrogen production, and residential energy consumption. To coordinate among such consumers, especially large, flexible loads like cloud computing facilities, we first need a standardized metric that defines the *real-time carbon intensity* of their electricity consumption. Here, *standardized* means a metric that is defined consistently across contexts, uses uniform units and calculation methods, is transparent and reproducible, and is widely adopted as a common reference so that results are comparable across different users, regions, and technologies. Such a metric should enable consumers to understand their carbon footprint from a system-wide perspective, and to adjust consumption patterns to meaningfully lower total system emissions.

However, lowering the carbon footprint of a consumer such as a data center does not always reduce overall grid emissions due to physical and technical complexity of power systems. Moreover, the choice of metric can strongly influence load-shifting strategies and their effectiveness. Therefore, selecting the right metric and control signal is critical.

## 3.1. Comparing Methods for Tracing Carbon Emissions

Multiple methodologies have been developed to quantify the carbon intensity of electricity consumption. Each method relies on specific assumptions regarding how emissions from generators should be attributed to electrical loads, and each exhibits distinct advantages and limitations. The following discussion outlines four commonly discussed approaches.

**1) Average carbon intensity:** This approach calculates carbon intensity by dividing the total emissions from all generators in a defined power system by the total electrical load over the same time period:

$$\text{ACI}(t) = \frac{E_{\text{total}}(t)}{L_{\text{total}}(t)}$$

where

$$E_{\text{total}}(t) = \sum_{g \in \mathcal{G}} e_g(t) \, P_g(t), \, L_{\text{total}}(t) = \sum_{i \in \mathcal{B}} L_i(t).$$

where $\mathcal{G}$ denote the set of generators (index $g$), $\mathcal{B}$ the set of loads (index $i$), and $t$ the time index. For each generator $g$ at time $t$, $P_g(t)$ is its electrical output and $e_g(t)$ its direct emission rate. For each load $i$, $L_i(t)$ is its electrical demand. This method is computationally straightforward, and ensures that the total emissions allocated to consumers match the measured system total. However, it assigns the same carbon intensity to all loads, disregarding transmission and distribution constraints and locational differences in generation. It also fails to capture the influence of load shifting on overall emissions and does not reflect unused low-carbon generation capacity, such as curtailed renewable energy.

**2) Flow-traced carbon intensity**: This approach extends average carbon accounting by tracing emissions from generating units to end-use loads using a proportional power-sharing principle, thus factoring in transmission constraints and locational differences in generation:

$$\text{FTCI}_i(t) = \sum_{g \in \mathcal{G}} f_{g,i}(t) \, e_g(t).$$

here $f_{g,i}(t) \in [0,1]$ is the fraction of generator $g$'s power that flows to load $i$ ($\sum_i f_{g,i}(t) = 1$), and $e_g(t)$ is its emission factor. Flow-traced signals are widely accessible through system operators [3] or third-party platforms such as Electricity Maps [4] and Singularity Energy [5].

**3) Locational marginal carbon intensity:** This approach estimates the short-term change in total system emissions resulting from a marginal change in load at a specific location:

$$\text{LMCI}_i(t) = \frac{\partial E_{\text{total}}(t)}{\partial L_i(t)}.$$

The derivative quantifies the incremental emissions that would result from a small load increase under the optimal-power-flow solution at time $t$. This approach estimates the short-term change in total system emissions resulting from a marginal change in load at a specific location. The value is typically derived from sensitivity analysis of the optimal power flow solution, which inherently incorporates transmission network constraints, but can also be inferred from a statistical analysis. While this method provides high locational resolution, verifying its real-life accuracy remains a challenge given that marginal changes cannot be empirically observed [6]-[7]. Furthermore, this approach reflects only the emissions of the marginal generator at that time and location, which may be either cleaner or more emitting than the system average. As a result, the total emissions allocated under this method diverge from the actual system-wide total.

**4) Adjusted locational marginal carbon accounting:** This method modifies the classical LMCI by adding a normalization term so that the emissions allocated to all loads exactly match the measured system total while preserving locational variation:

$$\text{ALMCI}_i(t) = \text{LMCI}_i(t) + \frac{E_{\text{total}}(t) - \sum_{j \in \mathcal{B}} L_j(t) \, \text{LMCI}_j(t)}{L_{\text{total}}(t)}$$

The additive adjustment term is uniform across buses and ensures $\sum_i L_i(t) \, \text{ALMCI}_i(t) = E_{\text{total}}(t)$. This method retains the locational specificity of the original method while improving consistency in system-level accounting. Nonetheless, it is generally applicable only for small changes in load and exhibits the same volatility driven by the emissions of the marginal generator, as the unadjusted version.

The advantages and drawbacks of these approaches are summarized in Table 1. Average carbon intensity offers simplicity and completeness in aggregate accounting but lacks the spatial and temporal resolution necessary for effective operational coordination. Flow-traced carbon intensity improves the spatial and temporal resolution of average emissions, by including transmission constraints. Locational marginal approaches improve locational accuracy but have to be verified, and are more appropriate for marginal adjustments in load rather than large-scale shifts. Greenhouse Gas Protocol has recently voted against pursuing marginal metrics for carbon accounting purposes, relying on short-term marginal signals may lead to higher emissions in the long run [8]. The selection of an appropriate method should therefore be guided by the intended application, the scale and nature of load adjustments, and the desired balance between accuracy and practical feasibility.

Although guidance for practitioners continues to evolve, most emerging standards favor the use of flow-traced emission factors. This preference reflects both their broader data availability and their stronger basis for transparent verification. The nodal information required to compute locational marginal metrics, such as detailed generation dispatch and curtailment data, is available only in a few regions and often with significant delays, whereas the inputs for average and flow-traced carbon intensities are consistently accessible worldwide. Moreover, locational marginal metrics rely on modeling a counterfactual system state that cannot be empirically verified.

**Table 1.** Comparison of four carbon accounting metrics used for real-time grid coordination, highlighting their definitions, advantages, and limitations.

| Carbon accounting metrics | Definition | Strengths | Limitations |
| --- | --- | --- | --- |
| Average | Total system emissions divided by total load | Simple, publicly available, ensures totals match | Ignores location and constraints, poor for load-shifting analysis, overlooks unused renewables |
| Flow-traced | Traces emissions from generators to loads using proportional flow assumptions | Direct conceptual allocation, includes transmission constraints | Not a quantification of immediate impact of load-shifting decision. |
| Locational marginal | Change in emissions from load change at a specific location | High locational accuracy, accounts for grid constraints | Consider only marginal generator, total may mismatch actual emissions, difficult to verify in practice |
| Adjusted locational marginal | Locational marginal adjusted so allocated totals equal system totals | Location-specific, fairer allocation | Small shifts only, volatile like locational marginal, difficult to verify in practice. |

## 3.2. Toward Dynamic, Privacy-Aware, and Fair Carbon Coordination Signals

The carbon intensity of grid electricity is inherently dynamic, varying not only across locations but also minute-to-minute as the generation fuel mix changes in response to demand, weather conditions, and operational constraints. As a result, a metric that is static or averaged over long periods may obscure opportunities for substantial emission reductions. While the carbon intensity metrics discussed above can provide useful insights, it is important to stress that annually-averaged versions will only enable *spatial load shifting*, that is, guiding where electricity consumption occurs within the network. High temporal resolution (e.g. hours of minutes) is required to make them suited for *temporal load shifting*, where the timing of consumption is adjusted.

To truly support system-wide decarbonization, we require more advanced carbon accounting frameworks that serve not only as equitable allocation tools for assigning emissions to assets, but also as operational signals for real-time coordination between consumers and the power system. Such signals should adapt to both spatial and temporal variations in carbon intensity, enabling flexible loads, such as data centers, industrial processes, and electric vehicle charging to shift operations in ways that reduce total system emissions, not merely the footprint of an individual site. This distinction is critical: the complexity of power system physics means that locally reducing emissions does not necessarily translate to a reduction in overall grid emissions, and in some cases, poorly coordinated load shifting could even increase them.

> Reducing emissions at an individual site does not guarantee, and may even hinder, a reduction in total grid emissions, if load shifting is poorly coordinated by not accounting for power system operational constraints.

Effective, physics-aligned carbon accounting depends on the availability of high-resolution, accurate, and timely data. Independent system operators and utilities should be encouraged, or required, to disclose real-time nodal carbon intensity data so that large consumers can align their operations with periods and locations of cleaner generation. However, increasing the spatial and temporal granularity of carbon data raises privacy concerns, especially when operational patterns or sensitive consumption behaviors of individual sites could be inferred from the data. These privacy considerations are particularly relevant for industrial facilities, commercial buildings, and residential consumers participating in demand-response programs. Coordinated carbon reduction efforts must therefore adopt privacy-preserving data sharing mechanisms, for example, through aggregation, anonymization, or the use of secure multi-party computation to ensure that actionable carbon signals can be shared without revealing confidential or commercially sensitive information.

While greater granularity generally improves the accuracy and usefulness of carbon signals, it also increases demands on data collection, processing, and governance. An adaptive granularity approach where the resolution of carbon intensity signals is tailored to the needs of specific applications, time scales, and privacy constraints offers a promising balance between effectiveness, feasibility, and confidentiality.

To maintain the credibility of coordination, several safeguards must be incorporated against "schedule washing," i.e., inflating baselines to claim artificial carbon savings. Reference schedules are derived from historical metered data and workload logs under comparable operating conditions. Incentives are tied to the difference between realized and baseline carbon-intensity exposure, using standardized metrics. Carbon-intensity data and emission factors are issued by grid operators or accredited third parties to ensure independence. Statistical monitoring detects and penalizes persistent upward drift in reported baselines.

Finally, the equity and fairness dimensions of carbon accounting remain paramount. Different allocation methodologies can assign markedly different levels of responsibility to various end-users, with direct social and economic implications. If these systems are designed without attention to socio-demographic differences, they may unintentionally shift disproportionate burdens onto communities with fewer resources to adapt or invest in cleaner alternatives. Embedding fairness as a design principle ensures that the costs and benefits of coordinated carbon reduction are distributed in a socially just manner.

> Future carbon intensity metrics should evolve from static, location-specific averages to **dynamic, privacy-aware, adaptive, fair, and system-aligned coordination signals**. Such signals will enable consumers and power systems to act in concert, maximizing the effectiveness of carbon reduction strategies while safeguarding both sensitive information and equitable outcomes.

## 3.3. AI-computing perspective on carbon tracing

While power-system methods define how emissions are allocated and aggregated, AI computing introduces additional challenges that complicate precise carbon tracing. First, AI workloads are geographically distributed and often migrate across data centers, making it difficult to associate compute activity with a single power node or region. Second, training and inference loads exhibit high temporal variability, unlike the relatively predictable demand of traditional consumers. Third, compute operations are largely managed by proprietary schedulers, creating a visibility gap between grid-level electricity use and application-level computing tasks. Finally, sharing high-resolution operational data can raise privacy and security concerns for both data-center operators and customers. These characteristics mean that accurate carbon accounting for AI computing requires bidirectional coordination: granular grid-level carbon signals must be paired with transparent, time- and location-aware reporting from AI workloads.

# 4. Green Scheduling and Planning

A carbon emissions metric only becomes meaningful when it moves beyond passive accounting and actively informs decision-making. To support deep decarbonization, carbon metrics must serve not just as retrospective reports but as real-time **coordination signals**, guiding when, where, and how electricity is consumed. In the context of power system scheduling and planning, this means providing actionable information to flexible consumers about the emissions consequences of their operations at specific times and locations. Carbon accounting addresses the question of *how much* is emitted and *by whom*; decision-making frameworks determine *what to do* in response. Bridging this gap is essential. Integrating carbon signals into the core of power system operations, across long-term expansion planning, short-term dispatch, real-time control, and market design, enables decisions that are not only cost-effective and reliable but also emissions-aware. In this section, we explore how carbon metrics can be embedded into scheduling and planning processes to drive coordinated, system-level emissions reductions across both centralized and decentralized actors.

## 4.1. Iterative Signal–response Coordination

In the iterative approach, the system operator runs a short, repeating cycle that aligns many small decisions with the system's emissions objective. At each cycle, the operator publishes a carbon coordination signal which is a time- and location-specific guidance that reflects current grid conditions and the emissions consequences of incremental load. Flexible consumers such as data centers adjust their schedules in response: deferring non-urgent jobs, advancing work into cleaner windows, migrating workloads across sites within latency limits, tuning model fidelity, to name but a few. Participants then return updated schedules, which the operator aggregates to refresh forecasts and recompute the signal for the next cycle. This closed loop repeats at the same interval the grid uses for redispatch in the operation phase (for example, every 5–15 minutes), with day-ahead previews to shape commitments. Iterative coordination largely prevents crowding, meaning everyone is rushing into the same "green" interval. The reason for that is after each round the operator recomputes the carbon signal using the updated aggregate load, so intervals that start to fill quickly become less attractive and demand shifts to neighboring periods. In practice the loop converges, but to avoid ping-pong swings when many actors move at once, we pair it with gentle smoothing and small, staggered starts, simple guardrails that make the adjustment gradual and the signal settle rather than oscillate. This architecture can be well-suited to day-ahead, intra-day and real-time operations, integrates cleanly with demand-response programs and ancillary service markets, and requires minimal assumptions about any one actor's internal logic. Its main challenges are tuning for stability during stressed conditions and maintaining high-quality forecasts.

## 4.2. Integrated Optimization with Learned Load Response

The integrated approach addresses the need for coherent day-ahead scheduling, market design, and long-horizon planning in environments with a few large, well-instrumented consumers. Instead of iterating in real time, the system operator first learns how major flexible loads typically respond to signals, prices, and constraints using historical data, field experiments, or contractual performance. The result is a response model that maps feasible load shifts to operational conditions and costs, with clear limits on timing, magnitude, latency, and reliability. These learned response functions are then embedded into a single optimization that co-optimizes grid operations and expected consumer adjustments in one solution. The formulation can represent consumer flexibility as piecewise envelopes, scenario-based response maps, or policy surrogates trained on operational data. Uncertainty is handled through robust or risk-aware terms so that plans remain conservative when behavior drifts. The solution produces consistent targets, tariffs, and schedules that already internalize realistic load shifting, curtailment absorption, and storage dispatch, thereby reducing the need for back-and-forth during execution. This integrated method excels at "shaping the field": setting carbon-indexed tariffs, valuing flexibility in procurement, designing emissions-aware market products, and informing siting and capacity expansion decisions. Results rely on how accurately the response model reflects real behavior, the integrated solves can be computationally intensive, and effective use requires data-sharing agreements. Regular retraining and periodic re-solves help track behavioral drift and incorporate new information. Note that, depending on the formulation, the integrated problem can be solved or updated fast enough for practical use, with linear and convex cases readily tractable and mixed-integer extensions handled through decomposition.

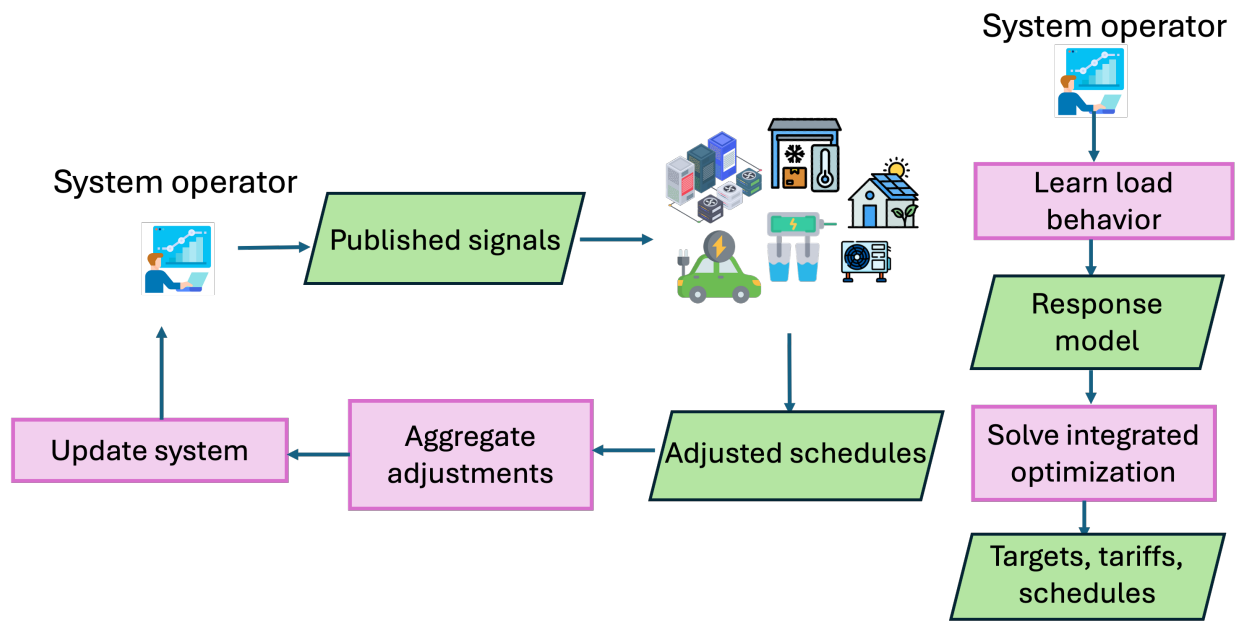

**Figure 2.** Two coordination methods for carbon-aware scheduling and planning. *Left:* Iterative signal–response loop where the operator publishes carbon signals; flexible loads adjust

schedules; aggregated adjustments update forecasts and grid state; the loop repeats until convergence. *Right:* Integrated optimization with learned load response where the operator learns typical flexibility offline, embeds that response in a single optimization, and produces targets, tariffs, and schedules. Operational data feedback to improve the response model. Rectangles show system functions or processes, while parallelograms indicate resulting outputs.

## 4.3. Reflecting on These Approaches

Figure 2 summarizes the two architectures developed above. The integrated optimization is used offline to design the rules of the game, signals, tariffs, targets, and infrastructure choices under realistic behavioral assumptions. The iterative loop then executes those rules online, correcting for forecast errors, outages, and human factors as they arise. Success of these approaches should be judged at the system level. That means proving real reductions in total emissions, cutting renewable curtailment, and improving clean-energy matching. It also means keeping reliability steady or better and ensuring that benefits and burdens are shared fairly. Yet several big questions remain: How can we quantify and certify counterfactuals for crediting without exposing sensitive data? How should carbon-indexed products be priced and enforced alongside existing energy and capacity markets? How do we prevent strategic behavior from actors who can influence forecasts? Further work on standardized, fast surrogate models and privacy-preserving and fair data sharing pipelines will accelerate adoption.

## 4.4. Uncertainty and Sensitivity

The present analysis assumes accurate carbon-intensity signals. In practice, these signals are subject to forecast and data uncertainty, and their impact on coordination outcomes has not yet been systematically quantified. Because both the iterative and integrated scheduling schemes rely mainly on the relative ranking of carbon signal values, moderate errors that do not alter this ordering are expected to have limited influence, whereas larger or correlated forecast errors could affect scheduling decisions. Assessing this sensitivity remains an important topic for future research as high-resolution carbon intensity forecasting continues to evolve.

## 4.5. Illustrative real-world applications

While large-scale coordination between data centers and power systems remains at an early stage, several emerging initiatives already demonstrate how the proposed scheduling principles can operate in practice. Examples include Google's 24/7 Carbon-Free Energy program, which shifts computing loads in alignment with hourly renewable availability [9]; National Grid ESO's publication of regional carbon-intensity data to inform demand-side scheduling [10]; and third-party platforms such as Electricity Maps [4] and Singularity Energy [5], which provide real-time flow-traced and marginal-emission signals. These examples illustrate the technical feasibility of carbon-aware scheduling and show that the proposed iterative and integrated frameworks are consistent with trends already taking shape in industry practice.

While this paper focuses on conceptual and architectural aspects, similar coordination frameworks have been examined quantitatively in recent studies. For example, study [11] applied a set of carbon-intensity signals, average, flow-traced, locational-marginal, and adjusted-locational, to the RTS-GMLC test system with flexible data-center loads. Their results show that the choice of signal strongly influences the temporal and spatial pattern of load shifting and the resulting system-level emissions, confirming the qualitative effects discussed here. These findings provide quantitative support for the coordination mechanisms and emission-reduction pathways described in this work.

# 5. Resilience

The coordination architectures introduced above do not only serve emission reduction. Under grid stress or emergencies, the same signal–response mechanisms can shift from "emissions-first" to "stability-first," enabling AI computing resources to support grid resilience and recovery. This continuity links sustainable operation and resilient operation within a single coordination framework. In normal times, the carbon signal steers consumption toward lower emissions. When the grid is under stress, that signal temporarily shifts to put stability and restoration first. The very actions used to reduce emissions, deferring non-urgent compute, migrating workloads across sites within latency and policy limits, dispatching on-site storage and thermal buffers, and adjusting model fidelity become tools to relieve congestion, support frequency and voltage, and speed recovery. As digital infrastructure becomes more critical, data centers and power systems are increasingly interdependent and coupled. Reliable electricity keeps compute available. Carbon-aware, flexible compute can, in turn, support the grid when conditions tighten. During crises, natural hazards, extreme weather, or large-scale outages, this synergy exposes vulnerabilities but also opens opportunities to strengthen system-level resilience. We examine here the following bidirectional relationship: how robust power systems can sustain data center operations during disruptions, and how data centers can actively contribute to power system resilience under stress.

## 5.1. Enhancing Data Center Resilience Through Power Systems

**Geographic Load Distribution and Grid-Aware Siting:** To minimize regional failure risk, many data centers adopt geo-redundant architectures, distributing workloads across multiple locations. Grid-aware placement factoring in local grid reliability, historical disaster frequency, and recovery speed enables intelligent rerouting of workloads during localized disruptions. This strategy is further strengthened through integration with microgrids and site-level renewables.

**Renewable Energy Integration with On-Site Storage:** Data centers increasingly leverage renewable generation coupled with battery systems or fuel-based generators to ensure continuity during outages. In microgrid or islanded modes, such hybrid configurations can operate autonomously, allowing facilities to sustain critical operations for hours or days, depending on storage capacity and workload prioritization.

**Demand Response Participation and Restoration Prioritization:** As critical infrastructure, data centers are often enrolled in demand response programs. In return for flexible energy consumption, utilities prioritize them during grid restoration. Through advanced coordination such as digital fault signaling or load curtailment protocols, data centers can proactively interface with utilities to accelerate recovery timelines.

### 5.2. Data Centers as Enablers of Power System Resilience

**Flexible Computational Workloads and Load Shedding:** Modern data centers possess the agility to shift, pause, or reschedule non-essential computing tasks in response to grid stress indicators like frequency drops or thermal overloads. Tasks such as deferred analytics, training of non-urgent AI models, or backup processing can be postponed or moved to other regions alleviating pressure on vulnerable local grids.

**Provision of Ancillary Grid Services:** With the integration of smart inverters and energy storage, data centers can function as distributed energy resources. They are capable of injecting reactive power, participating in frequency regulation, and providing spinning reserves. Some advanced architectures transform data centers into virtual power plants, orchestrating distributed contributions to stabilize the grid during peak load or emergency conditions.

**Blackstart Support and Global Load Redistribution:** In grid collapse scenarios, data centers with local generation capabilities can support blackstart efforts, jump-starting segments of the grid by providing stable voltage and frequency references. Furthermore, global scheduling platforms can shift workloads across continents, effectively redistributing compute demands away from distressed grids while maintaining service availability.

# Further Readings

# Biographies

Farzaneh Pourahmadi is with the Technical University of Denmark, Department of Wind and Energy Systems, 2800 Kgs Lyngby, Denmark

Olivier Corradi is with Electricity Maps, 2200 Copenhagen, Denmark

Pierre Pinson is with Imperial College London, Dyson School of Design Engineering, London SW7 2AZ, United Kingdom